\documentclass[twocolumn,showpacs,amsmath,amssymb,aps]{revtex4}   %%  4-pages !!
\usepackage{graphicx}% Include figure files
\usepackage{dcolumn}% Align table columns on decimal point
\usepackage{bm}% bold math
\usepackage{color}% color text
\usepackage{amsmath}

\renewcommand{\vec}[1]{\mathbf{#1}}

\newif\ifgraph

\graphtrue             %
\begin{document}
\title{
Non-Gaussian Normal Diffusion in Low Dimensional Systems}

\author{Qingqing Yin$^{1}$, Yunyun Li$^{1, \dag}$,  Fabio Marchesoni$^{1,2,\dag}$, Shubhadip Nayak$^{3}$, and Pulak K. Ghosh$^{3 \dag}$}

 \affiliation{Center for Phononics and Thermal Energy Science, Shanghai Key Laboratory of Special Artificial Microstructure Materials and Technology, School of Physics Science and Engineering, Tongji University, Shanghai 200092, China\\
Dipartimento di Fisica, Universit\`{a} di Camerino, I-62032 Camerino, Italy\\
Department of Chemistry, Presidency University, Kolkata 700073, India\\
Corresponding authors.\ E-mail: $^ \dag$yunyunli@tongji.edu.cn, $^\dag$ fabio.marchesoni@pg.infn.it, $^\dag$ pulak.chem@presiuniv.ac.in}

\date{\today}

\begin{abstract}
 Brownian particles suspended in disordered crowded environments often
exhibit non-Gaussian normal diffusion (NGND), whereby their displacements
grow with mean square proportional to the observation time and non-Gaussian
statistics. Their distributions appear to decay almost exponentially according to ``universal'' laws largely insensitive to the observation
time. This effect is generically attributed to slow environmental
fluctuations, which perturb the local configuration of the suspension
medium. To investigate the microscopic mechanisms responsible for the NGND
phenomenon, we study Brownian diffusion in low dimensional systems, like
the free diffusion of ellipsoidal and active particles, the diffusion of
colloidal particles in fluctuating corrugated channels and Brownian motion
in arrays of planar convective rolls. NGND appears to be a transient effect
related to the time modulation of the instantaneous particle's diffusivity,
which can occur even under equilibrium conditions. Consequently, we propose
to generalize the definition of NGND to include transient displacement
distributions which vary continuously with the observation time. To this
purpose, we provide a heuristic one-parameter function, which fits all
time-dependent transient displacement distributions corresponding to the
same diffusion constant. Moreover, we reveal the existence of low
dimensional systems where the NGND distributions are not leptokurtic (fat
exponential tails), as often reported in the literature, but platykurtic
(thin sub-Gaussian tails), i.e., with negative excess kurtosis. The actual
nature of the NGND transients is related to the specific microscopic
dynamics of the diffusing particle. 
\end{abstract}

\maketitle

\section{Introduction} \label{intro}
Possibly misinterpreting the original works of Albert Einstein and Marian
Smoluchowski on Brownian motion, one tends to associate the normal diffusion
of an ideal Brownian particle with the Gaussian distribution of its spatial
displacements. Recent observations
\cite{Granick1,Granick2,Bhatta,Sung1,Sung2,Granick3} of Brownian motion in
fluctuating crowded environments led to question the generality of this
notion. Indeed, it implicitly assumes Fick's diffusion \cite{Gardiner},
whereby the directed displacements of an overdamped particle, say, in the $x$
direction, $\Delta x(t)=x(t)-x(0)$, would grow according to the asymptotic
Einstein law, $\langle \Delta x^2(t) \rangle = 2Dt$, and with Gaussian
statistics. The probability density function (pdf) of the rescaled
observable, $\delta_t=\Delta x/\sqrt{t}$, would thus be a stationary Gaussian
function with half-variance $D$.

However, there are no fundamental reasons why the diffusion of a physical
Brownian tracer should be of the Fickian type. For instance, in real
biophysical systems, displacement pdf's have been reported, which retain
prominent exponential tails over extended intervals of the observation time,
even after the tracer has attained the asymptotic condition of normal
diffusion. Such an effect, often termed non-Gaussian normal diffusion (NGND),
disappears only for exceedingly long observation times (possibly inaccessible
to real experiments \cite{Granick1}), when the displacement distributions
eventually turn Gaussian, as dictated by the central limit theorem, without
changes of the  diffusion constant. Persistent diffusive transients of this
type have been detected in diverse experimental setups
\cite{Granick1,Granick2,Bhatta,other1,other3,other4}. Extensive numerical
simulations confirmed the occurrence of NGND in crowded environments
featuring slowly diffusing or changing microscopic constituents (filaments
\cite{Granick1, Bhatta}, large hard spheres \cite{Sung1,Sung2,Granick3},
clusters \cite{Kegel,Kob}, and other heterogeneities \cite{Tong,Cherstvy}).

The current interpretation of this phenomenon postulates the existence of one
or more fluctuating processes affecting composition and geometry of the
particle's suspension medium \cite{Granick1}. It seems reasonable that, for
observation times comparable with the relevant environmental relaxation
time(s), the tracer displacements may obey a non-Gaussian statistics.
 The rescaled  pdf's, $p(\delta_t)$, are
expected to be Gaussian for both much shorter and much larger observation
times, but with different half-variance: the free diffusion constant, $D_0$,
for $t \to 0$ (no crowding effect) and the asymptotic diffusion constant,
$D$, introduced above, for $t \to \infty$ (central limit theorem). The
mechanism how the tracer's normal diffusion sets in and the constant $D$
remains unaltered through the entire non-Gaussian transient, varies, instead,
from case to case. In summary, key features of the NGND phenomenon appear to
be: (i) its transient nature, whereby the observables taken into account are
intrinsically non-stationary; (ii) a time-modulated instantaneous diffusivity
of the tracer. As discussed in the following, these conditions can occur even
in the absence of external (non-equilibrium) perturbations of the Brownian
dynamics.

A simple heuristic explanation \cite{Granick1} of the NGND phenomenon models
the effects of the slowly fluctuating environment in terms of an {\it ad hoc}
distribution of the tracer's diffusion constant. Imposing an exponential
distribution of the diffusion constant with average $D$, a straightforward
superstatistical procedure yields the exponential (Laplace) rescaled
distribution, $p(\delta_t) = \exp(-\delta_t/\alpha)/2\alpha$, with $\alpha^2
= D$. A more suggestive NGND paradigm is provided by the notion of {\it
diffusing diffusivity} \cite{Slater}, whereby the asymptotic particle's
diffusion constant is replaced by a time fluctuating auxiliary observable,
$D(t)$. Regarding $D(t)$ as a continuous stochastic process with average $D$
and time constant $\tau$, the distribution $p(\delta_t)$ changes from
exponential for $t \ll \tau$ to Gaussian for $t\gg \tau$; in both time
regimes, the displacement diffusion is normal, with $\langle \Delta x^2(t)
\rangle = 2Dt$ \cite{Slater}. Refined variations of these paradigms
\cite{Metzler2,Jain1,Jain2,Tyagi,Luo,Sokolov,Metzler}, predict different
exponential decays of the transient distributions. These phenomenological
approaches have two major limitations, namely: (i) They fail to incorporate
the free Gaussian diffusion detected in most real and numerical experiments
at very short observation times, $t \to 0$, when crowding plays no role. This
is because these approaches purportedly ignore the microscopic details of the
actual diffusion mechanisms;
(ii) %with the significant exception of Ref. \cite{Sokolov},
They generally aim at ``universal'' non-Gaussian transient pdf's, namely, at
functions $p(\delta_t)$ insensitive to $t$ over extended domains, $t < \tau$. [In the superstatistical models $\tau = \infty$.] However, even if
this strategy may appear to agree with NGND observations for complex systems
\cite{Granick1,Granick2,Bhatta,Sung1,Sung2,Granick3}, it is obvious that to
reproduce the exponential-Gaussian crossover, the transient rescaled
distributions must assume the form $p(\delta_t, t)$, i.e., they must depend
explicitly on $t$.

This study focuses on {\em the microscopic mechanisms responsible for NGND}.
To this purpose, motivated by a preliminary study \cite{PRR}, we
investigated, both numerically and analytically, directed diffusion of
different idealized tracers in confined geometries. We selected low
dimensional systems mostly inspired to cell biology \cite{RMP_BM,Lipowski}.
For appropriately short observation times, NGND emerges as a transient effect
of the time modulation of the tracers' microscopic diffusivity. This effect
can occur even in the absence of environmental fluctuations. It suffices to
require that the tracer's dynamics be governed by two concurring diffusion
mechanisms, at least one of them characterized by a finite relaxation time,
$\tau$. During transients times of the order of $\tau$, the displacement
distributions can deviate from their asymptotic Gaussian profile also after
normal diffusion has set in. Moreover, such deviations do not necessarily
imply the emergence of ``fatter'' exponential tails (leptokurtic transients),
but under certain conditions, the distribution tails can get ``thinner''
(platykurtic transients).

This observation suggests typical NGND features are to be found in much wider
a class of diffusion systems. Indeed, contrary to experimental and numerical
observations on extended systems, the NGND transient displacement
distributions in low dimensional models, are found to depend on the
observation time. This led us to address the question of phenomenological
fitting functions capable of reproducing the $t$-dependence of the rescaled
pdf's, $p(\delta_t,t)$.  We also noticed that the $t$ dependence of the
transient rescaled pdf's can be suppressed, though not completely, by
considering models where the onset of normal diffusion is controlled by some
intrinsic time constant, which can be taken much shorter than the upper
bound, $\tau$, of the non-Gaussian transient. This provides us with a
criterion to formulate low dimensional models that better capture the known
NGND phenomenology in complex systems.

The present paper is organized as follows. In Sec. \ref{model1} we elaborate
on a toy discrete model of NGND  proposed first in Ref. \cite{Slater} and
then revisited in Ref. \cite{PRR}. The purpose of this section is to single
out key NGND aspects, like the time scales regulating the diffusion
mechanisms and the nature of the non-Gaussian transients, that is, lepto-
versus platykurtic. In Sec. \ref{Perrin} we consider the diffusion of a two
dimensional (2D) ellipsoidal Brownian particle in a highly viscous,
homogeneous and isotropic fluid in thermal equilibrium. For observation times
shorter than its rotational relaxation time, the particle does undergo normal
diffusion. However, its instantaneous diffusivity in a given direction is
modulated in time due to its elongated shape. This results in exponentially
decaying transient distributions of the particle's directional displacements.
In Sec. \ref{Janus} we analyze the diffusion of a 2D self-propelling
symmetric particle in a homogeneous and isotropic active medium with finite
orientational relaxation time. NGND is characterized here by thin tails of
the transient displacement distributions. In both cases, however, transients
are governed by one time scale, only, their orientational diffusion time,
$\tau$: on increasing the observation time, normal diffusion just anticipates
the onset of the Gaussian statistics of the particle's displacements. In Sec.
\ref{beta} we introduce a phenomenological fitting function,
$p_\beta(\delta_t)$ for the rescaled displacement distributions, with only
one adjustable parameter, $\beta$. This function is designed {\em ad hoc} to
ensure normal diffusion with the observed diffusion constant, $D$, at any
time, while $\beta$ encodes the $t$-dependence of the rescaled displacement
distributions. Relevant values of the fitting parameter $\beta$ are $\beta=2$
for a Gaussian pdf, $\beta=1$ for a Laplace (exponential) pdf, $\beta<2$
for a leptokurtic pdf, and $\beta>2$ for a platykurtic pdf. In Sec. \ref{channel} we analyze the NGND phenomenon
in a narrow corrugated channel \cite{chemphyschem,PNAS} with fluctuating
pores \cite{PRR}. NGND occurs for time-correlated pore fluctuations, random
and periodic, alike, and, more importantly, for observation times comprised
between two distinct, controllable time scales. The correlation time of the
pore fluctuations sets the transient time scale, $\tau$, whereas the average
pore-crossing time governs the onset of normal diffusion. Upon choosing the
former much larger than the latter, the NGND transient is made grow wider and
the $t$-dependence of $\beta$ weaker.  As a practical application of the
tools introduced thus far, in Sec. \ref{Pomeau} we investigate the diffusion
of a passive Brownian tracer in a periodic array of planar counter-rotating
convection rolls. The peculiarity of this model is that, by tuning its
dynamical parameters, transients can change from lepto- to platykurtic. Two
are the systems's characteristic time scales: the mean time for the particle
to first exit a convection roll and its average revolution period inside the
roll. At low (high) temperatures, the former (latter) time scale is larger
and thus plays the role of transient time, $\tau$; accordingly, the NGND
transients are leptokurtic (platykurtic). Finally, in Sec. \ref{conclusions}
we summarize the main conclusions of our approach to NGND.
% and discuss its applications to other physical systems.

\section{A Discrete NGND Model} \label{model1}

Though sounding exotic to some readers, the phenomenon of NGND turns out to
be way more general than the more familiar Fickian diffusion. To make this
point, we elaborate now on a coarse grained model, first proposed in Ref.
\cite{Slater}, which serves well the purpose of illustrating NGND in
continuous systems of any dimensionality.

Let us coarse grain the trajectory of a tagged particle in the $x$ direction
as the sum of small random steps, $\Delta x_i$, taken at fixed discrete
times, $t_i=i\Delta t$, where $i=1, \dots N$ and $\Delta t=1$, for
simplicity. Accordingly, the position of the particle at time $N$ is
$x_N=\sum_{i=1}^{N} \Delta x_i$. A stochastic average over the particle's
steps $\Delta x_i$ yields the  mean square displacement at time $N$,
\begin{equation} \label {sl1}
\langle x_N^2\rangle= \sum_{i=1}^{N}\langle \Delta x_i^2\rangle +
2\sum_{i \neq j}{}^\prime \langle \Delta x_i \Delta x_j\rangle,
\end{equation}
where $\sum_{i \neq j}^\prime $ stays for $\sum_{i=1}^{N-1}\sum_{j=i+1}^{N}$.
Sufficient conditions to establish normal diffusion are that: (1) the {\it
step directions} are uncorrelated, $\langle \Delta x_i \Delta x_j \rangle=0$,
that is, for any given $\Delta x_i$, displacements $\Delta x_j$ and $-\Delta
x_j$, are equiprobable; (2) the variance, $\langle \Delta x_i^2 \rangle$, of
the step probabilities, $p(\Delta x_i)$, are of the same order of magnitude,
though not necessarily identical. These requirements are less stringent than
the assumptions implicit in the standard random walker model for Brownian
motion \cite{Gardiner}.

Indeed, for the sake of generality, one should not rule out finite
correlations of the {\it step lengths} \cite{Slater}. For instance, we can
assume that during each unit time step the particle's diffusion is normal
with time-dependent constant, $D_i$, i.e.,
\begin{equation} \label{slpD}
p(\Delta x_i) = (4\pi D_i)^{-1/2} \exp(-\Delta x_i^2/4D_i).
\end{equation}
This assumption guarantees that the directions of the particle's steps are
uncorrelated, while their length correlation is controlled by the
auto-correlation of the time sequence of the constants $D_i$, which, in turn,
is specific to the system at hand. It follows immediately that
\begin{equation}\label{sl2}
\langle x_N^2\rangle = 2\langle D\rangle N,
\end{equation}
and
\begin{eqnarray}
\langle x_N^4\rangle - 3\langle x_N^2\rangle^2= 12\mu_D\langle D \rangle^2 N+24{\sum_{i\neq j}}^\prime C_{ij},\label{sl4}
\end{eqnarray}
with $\mu_D= (\langle D^2\rangle-\langle D\rangle^2)/\langle D\rangle^2$ and
$C_{ij}=\langle D_i D_j\rangle-\langle D_i\rangle \langle D_j\rangle$. For
any given stationary model, there exists an appropriate distribution of the
constants $D_i$, $p(D_i)$, so that $\langle D_i \rangle \equiv \langle D
\rangle$.

Suppose now that two particle steps, $\Delta x_i$ and $\Delta x_j$ are
statistically uncorrelated only for large time differences, i.e., $\langle D_i D_j\rangle=\langle D_i
\rangle\langle D_j \rangle$ for
$|i-j|> \tau$. We then distinguish two limiting cases,

(i) $N \gg \tau$, where
\begin{equation}\label{slG}
\mu_x=\frac{\langle x_N^4\rangle - 3\langle x_N^2\rangle^2}{\langle x_N^2\rangle^2} = \frac{3 \mu_D}{N} \rightarrow 0.
\end{equation}
A vanishing excess kurtosis, $\mu_{x}$, hints at a Gaussian $x_N$
distribution. This is the asymptotic limit of the displacement distributions
predicted by the central limit theorem.

(ii) $N < \tau$, where
\begin{equation}\label{slL}
  \mu_x=\frac{\langle x_N^4\rangle - 3\langle x_N^2\rangle^2}{\langle x_N^2\rangle^2}  = 3 \bar \mu_D.
\end{equation}
with $\bar \mu_D=(2/N^2\langle D^2 \rangle)\sum_{i \neq j}{}^\prime C_{ij}$.
Eqs. (\ref{sl2}) and (\ref{slL}) embody the definition of NGND. The finite
excess kurtosis, $\mu_x$ depends on the actual auto-correlation of the
constants $D_i$. For instance, on assuming $\langle D_i D_j\rangle=\langle
D^2\rangle$ for all $i$ and $j$ with $|i-j|<\tau$, we obtain $\bar
\mu_D=\mu_D$. In particular, for the exponential distribution
$p(D_i)=\exp(-D_i/\langle D \rangle)/\langle D \rangle$ assumed in the
diffusing diffusivity model of Ref. \cite{Slater}, $\mu_D=1$. Not
surprisingly, the resulting value of the excess kurtosis, $\mu_x=3$,
corresponds to a Laplace distribution of the total displacement $x_N$
\cite{Slater}.

Of course a more realistic choice of the correlator $C_{ij}$ can yield
different values of $\mu_x$. In most applications $C_{ij}$ is definite
positive and decays to zero with time, i.e., with $|i-j|$; hence $0 < \mu_x
<3$, Accordingly, the corresponding $x_N$ distributions are leptokurtic, with
tails decaying slower than those of a Gaussian distribution, but typically
faster than exponentially. On the other hand, we cannot exclude the
possibility that $C_{ij}$ decays to zero oscillating. This implies that, in
principle, $\mu_D$ can assume negative values, so that the corresponding
transient distribution of $x_N$ may be platykurtic. In Refs.
\cite{nonG1,nonG2,nonG3} the present approach has been extended also to
microscopically non-Gaussian diffusive processes [where the $\Delta x$
distribution of Eq.(\ref{slpD}) does not apply].

We conclude this section with a final remark about the time scales involved
in this discrete model. One time scale has been introduced explicitly, namely
the characteristic decay time, $\tau$, of the correlator $C_{ij}$ or,
equivalently, the correlation time of the step lengths, $\Delta x_i$. A
second one is implicit in our choice for the step distribution, $p(\Delta
x_i)$. In Eq. (\ref{slpD}) the coarse grained diffusion was assumed to be
normal over the time step $\Delta t=1$. This implies that in the
corresponding continuum system normal diffusion is expected to have occurred
at some intrinsic time scale much shorter than $\tau$. Of course the discrete
model of this section cannot reproduce the diffusion properties at times
shorter than the discretization time scale, $\Delta t$.

\section{Diffusion of an Ellipsoidal Particle} \label{Perrin}

\begin{figure}[t!]
\centering \includegraphics[width=8.5cm]{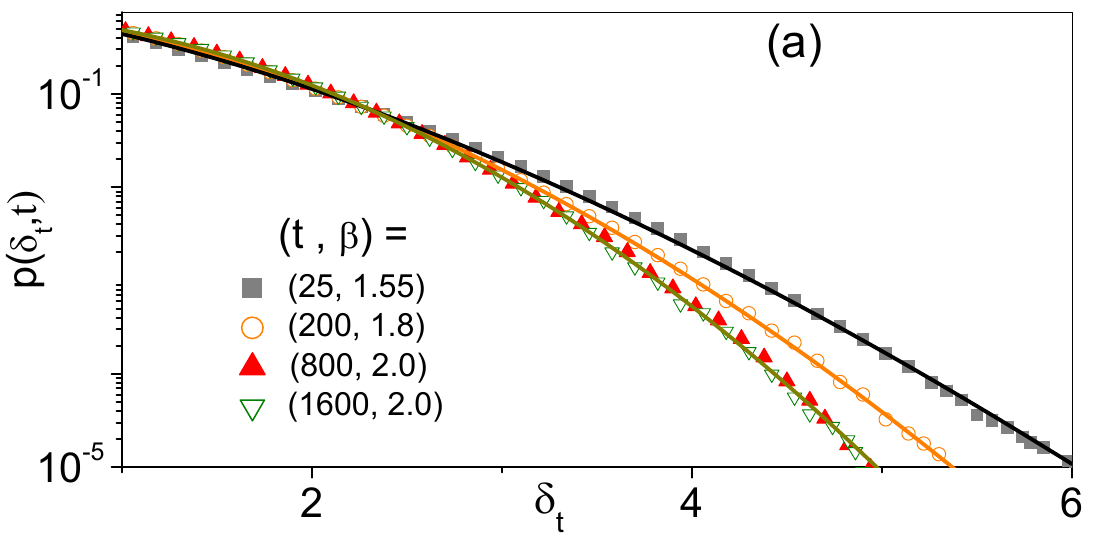}
\centering \includegraphics[width=8.5cm]{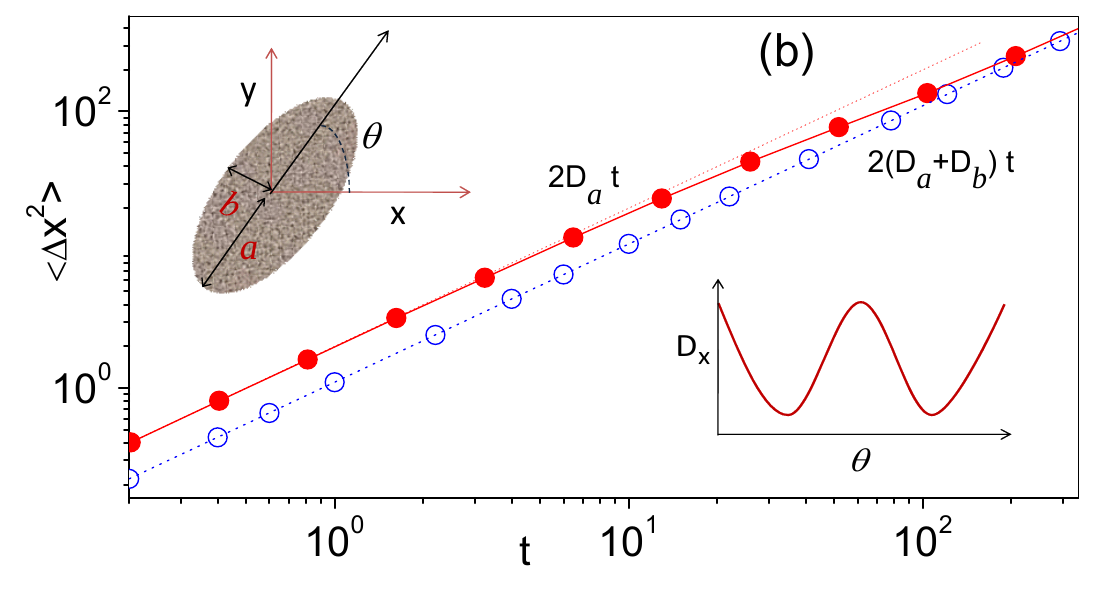}
\centering \includegraphics[width=8.5cm]{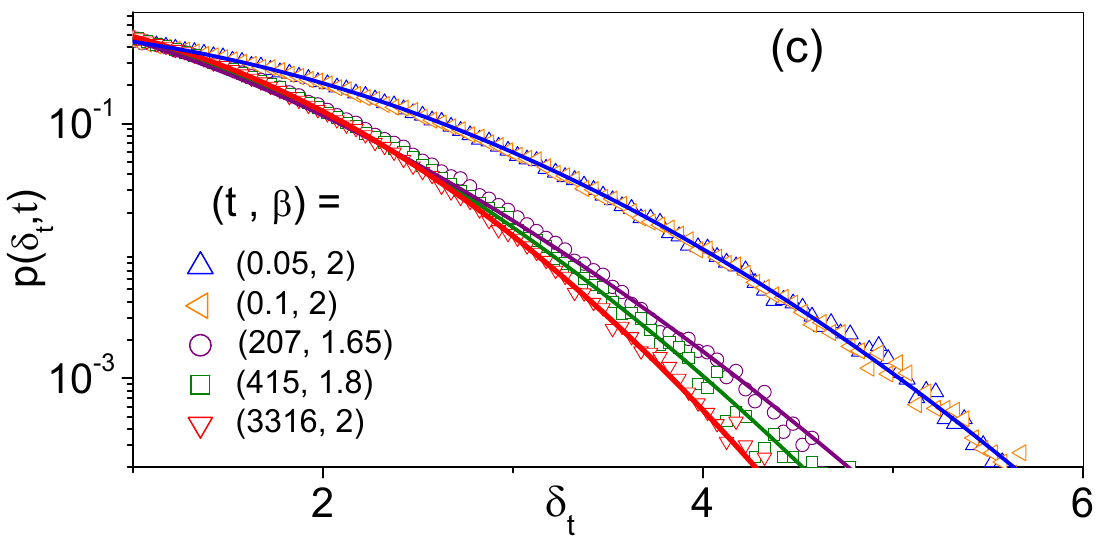}
\caption{Overdamped 2D ellipsoidal particle of semi-axes $a=0.5$ and $b=0.05$
diffusing in a homogeneous medium with Langevin Eqs. (\ref{P1})-(\ref{P3}):
(a), (c) displacement pdf's for different initial orientations [uniform $\theta(0)$
distribution in (a), and $\theta(0)=0$ in (c)] and increasing observation
times, $t$, see legends; (b)
$\langle \Delta x^2\rangle$ vs. $t$ for the initial conditions (i.c.) of
(a) (empty symbols) and (c) (filled symbols).
Simulation parameters are: $D_a=1$, $D_b=(b/a)D_a$ and $D_r=0.01$.
Asymptotic diffusion in (b) follows the normal diffusion law, $2Dt$ with
$D=(D_a+D_b)/2$ (dashed line), independent of the i.c. At very short times, the
diffusion constant depends on $\theta(0)$ (see sketch).
The pdf's  have been fitted by means of Eq. (\ref{B3}) for
$D$ fitting the large-$t$ simulation data of (b) and $\beta$ as reported in the legends.
\label{F1}}
\end{figure}

We consider first the simple case of a 2D ellipsoidal particle of semiaxes
$a$ and $b$, with $a > b$, diffusing in a highly viscous, homogeneous and
isotropic medium, subject to equilibrium thermal fluctuations. This is a
well-known problem in biological physics \cite{Berg}. The particle's
elongation causes a dissipative coupling between the center of mass
translational degrees of freedom, $x$ and $y$ in the laboratory frame, and
the rotational degree of freedom, $\theta$. As sketched in Fig. \ref{F1}(b),
the angle $\theta$ defines the orientation of the particle's long axis with
respect to the horizontal $x$ axis. The physical consequences of such a
mechanism were first recognized by F. Perrin \cite{Perrin}. An ellipsoidal
particle tends to diffuse independently in directions parallel and
perpendicular to its long axis, that is along its principal axes. The
relevant diffusion constants in the body frame are denoted here by $D_a$ and
$D_b$, with $D_a \geq D_b$. In 2D, rotational diffusion is governed by an
additional diffusion constant, $D_\theta$, which will be handled here as
unrelated to the translational constants, $D_a$ and $D_b$, to avoid
unnecessary complications involving hydrodynamic effects and fabrication
issues \cite{Berg,Perrin}. Over the angular relaxation time
$\tau=1/D_\theta$, random diffusion erases any directional memory of the
particle's motion. Related to this mechanism is the crossover between
anisotropic diffusion with constants $D_a$ and $D_b$ at short observation
times, $t \ll \tau$, and isotropic diffusion with constant $D=(D_a+D_b)/2$ at
long observation times, $t>\tau$ \cite{ellipsoid}.

The anisotropic-isotropic crossover can be numerically investigated by
integrating the Langevin equations \cite{Kloeden} describing the
roto-translational motion of a free ellipsoidal Brownian particle,
\begin{eqnarray}
\dot x&=& \xi_x(t), ~~~ \dot y=\xi_y(t), \label {P1}\\
\dot \theta&=&\xi_\theta (t) \label{P2},
\end{eqnarray}
where the translational noises, $\xi_i(t)$ with $i=x,y$, and the rotational
noise, $\xi_\theta (t)$, model three independent stationary Gaussian
fluctuation sources with zero means and autocorrelation functions $\langle
\xi_i(t) \xi_j(0)\rangle=2D_{ij} \delta(t)$ and $\langle
\xi_\theta (t) \xi_\theta (0) \rangle = 2 D_\theta \delta (t)$.
The matrix $D_{ij}$ encodes the
dissipative roto-translational coupling, namely \cite{ellipsoid}
\begin{equation} \label{P3}
D_{ij}= ({1}/{2})[(D_a+D_b)\delta_{ij}+(D_a-D_b)M_{ij}(\theta)],
\end{equation}
with ${\bf M}=\begin{pmatrix}
\cos 2\theta & \sin 2\theta \\
\sin 2\theta & -\cos 2\theta
\end{pmatrix}$.

Despite their apparent simplicity, the analytical solution of the Langevin
Eqs. (\ref{P1})-(\ref{P3}) is rather cumbersome \cite{Pecora,Franosch}. The
diffusion properties of a typical ellipsoidal particle are summarized in Fig.
\ref{F1}. The mean square displacement, $\langle \Delta x^2(t)\rangle$,
plotted in panel (b) as a function of the observation time $t$, was first
computed under assuming a uniform distribution of $\theta(0)$. This initial
condition (i.c.) was justified with the practical difficulty of measuring the
particle instantaneous orientation and with the isotropy of the suspension
medium. The resulting asymptotic diffusion constant, numerically determined
as
$$D=\lim_{t\to\infty}\langle \Delta x^2(t)\rangle/2t,$$ agrees with the
expected value, $D=(D_a+D_b)/2$, obtained by averaging $D_{ij}(\theta)$ in
Eq. (\ref{P3}) with respect to the isotropic equilibrium distribution of
$\theta$. This result is indeed an effect of our choice for the i.c. of
$\theta$. A stochastic average over a uniform $\theta(0)$ distribution is
equivalent to imposing isotropic particle's diffusion, that is establishing
Einstein law at any time. In contrast, by setting $\theta(0)=0$, the
numerical data for $\langle \Delta x^2\rangle$ versus $t$, also shown in Fig.
\ref{F1}(b), bridge two linear laws with different diffusion constant:
$D=D_a$ for $t\ll \tau$ and $D=(D_a+D_b)/2$ for $t \gg \tau$.

More revealing are the distributions of the unidirectional displacements,
$\Delta x$, for increasing observation times, $t$, plotted in panels (a) and
(c), respectively for a uniform initial angular distribution and
$\theta(0)=0$. As first theoretically predicted by Prager \cite{Prager} and
numerically confirmed by the authors of Ref. \cite{Franosch}, for both i.c.
the rescaled displacement pdf's do approach the Gaussian profile of Fickian
diffusion with half-variance $D=(D_a+D_b)/2$, but only for $t \gg \tau$, that
is well after the anisotropic-isotropic crossover took place. Most
remarkably, for $\theta(0)=0$, in panel (c),  the displacement distributions
approach a Gaussian profile both for $t \ll \tau$ and $t \gg \tau$, each with
the corresponding half-variance $D$ shown in panel (b), respectively, $D_a$
and $(D_a+D_b)/2$. The short-$t$ ``reentrant'' Gaussian distribution does not
appear in panel (a), due to the randomized i.c.. The explanation of this
behavior is simple. In panel (c), the particle's long axis was initially
oriented parallel to the $x$ axis, $\theta(0)=0$. Therefore, it started
diffusing in the $x$ direction like a one dimensional Brownian particle, with
diffusion constant $D_a$. Subsequently, angular fluctuations mixed diffusion
along the two symmetry axes with time constant $\tau$. This argument can be
extended to any choice of $\theta(0)$: based on Eq. (\ref{P3}), the short-$t$
diffusion constant is expected to be $D=(1/2)[(D_a+D_b)+(D_a-D_b)\cos
\theta(0)]$, see inset of Fig. \ref{F1}(b). Of course, the i.c. only
influence the anisotropic diffusion regime at short $t$.

There is only one characteristic time scale in this model, namely, the
angular relaxation time, $\tau=1/D_\theta$. However, normal diffusion turns
out to set in for shorter observation times, $t \sim \tau$, than the
displacement Gaussian statistics. To explain this behavior, we notice that
during the transient time, $\tau$, a maximum mean square displacement, $2D_a
\tau$, occurs parallel to the major axis; observing the same displacement in
the perpendicular direction would take a larger time, $\tau_*=\tau(D_a/D_b)$.
The onset of the Gaussian $\Delta x$ statistics is thus delayed to larger
observation times with $t > \tau_*$.

In conclusion, this simple model of equilibrium Brownian motion exhibits
NGND. On decreasing the observation time, $t$, the rescaled displacement
distribution in a fixed laboratory direction, changes from Gaussian for $t
\gg \tau$, to a leptokurtic distribution with fat exponential tails for $t
\sim \tau$, independently of the i.c.. This behavior is consistent with the
phenomenological picture of Sec. \ref{model1}. This is apparent in the case
of uniform initial orientation. Normal diffusion is ensured by the fact that,
after the particle has taken a step $\Delta x_i$ at the discrete time
$t_i=i\Delta x$, it will next take a step $\pm \Delta x_j$ at time $t_j$,
with equal probability. On the contrary, the step lengths $\Delta x_i$ and
$\Delta x_j$ are correlated for $|t_j-t_i| < \tau$. Indeed, the
effective half-width of the diffusing particle parallel to the $x$ axis,
varies randomly between $b$ at $\theta =0,\pi$ and $a$ at $\theta=\pm \pi/2$.
Accordingly, the particle's instantaneous diffusion constant fluctuates
between $D_a$ and $D_b$; its fluctuations are exponentially time correlated
with time constant $\tau$. As discussed in Sec. II, this leads to a rescaled
pdf, $p(\delta_t,t)$, with positive excess kurtosis.

\section{Diffusion of a Janus Particle} \label{Janus}

We consider next the case of a pointlike particle undergoing persistent
Brownian motion, namely, a 2D artificial microswimmer. Typical artificial
microswimmers are Brownian particles capable of self-propulsion in an active
medium \cite{Granick,Muller}. Like in the foregoing section, the suspension
medium can be taken homogeneous, isotropic and highly viscous. Such particles
are designed to harvest environmental energy by converting it into kinetic
energy. The simplest class of artificial swimmers investigated in the
literature are the so-called Janus particles (JP), mostly spherical colloidal
particles with two differently coated hemispheres, or ``faces''
\cite{Marchetti,Gompper}. Recently, artificial micro- and nanoswimmers of
this class have been the focus of pharmaceutical (e.g., smart drug delivery
\cite{smart}) and medical research (e.g., robotic microsurgery \cite{Wang}).
Relevant to the present work is the observation that their function is
governed, in time and space, by their diffusive properties through complex
environments, which are often spatially patterned \cite{Bechinger} or
confined \cite{ourPRL}.
\begin{figure}[t!]
\centering \includegraphics[width=8.5cm]{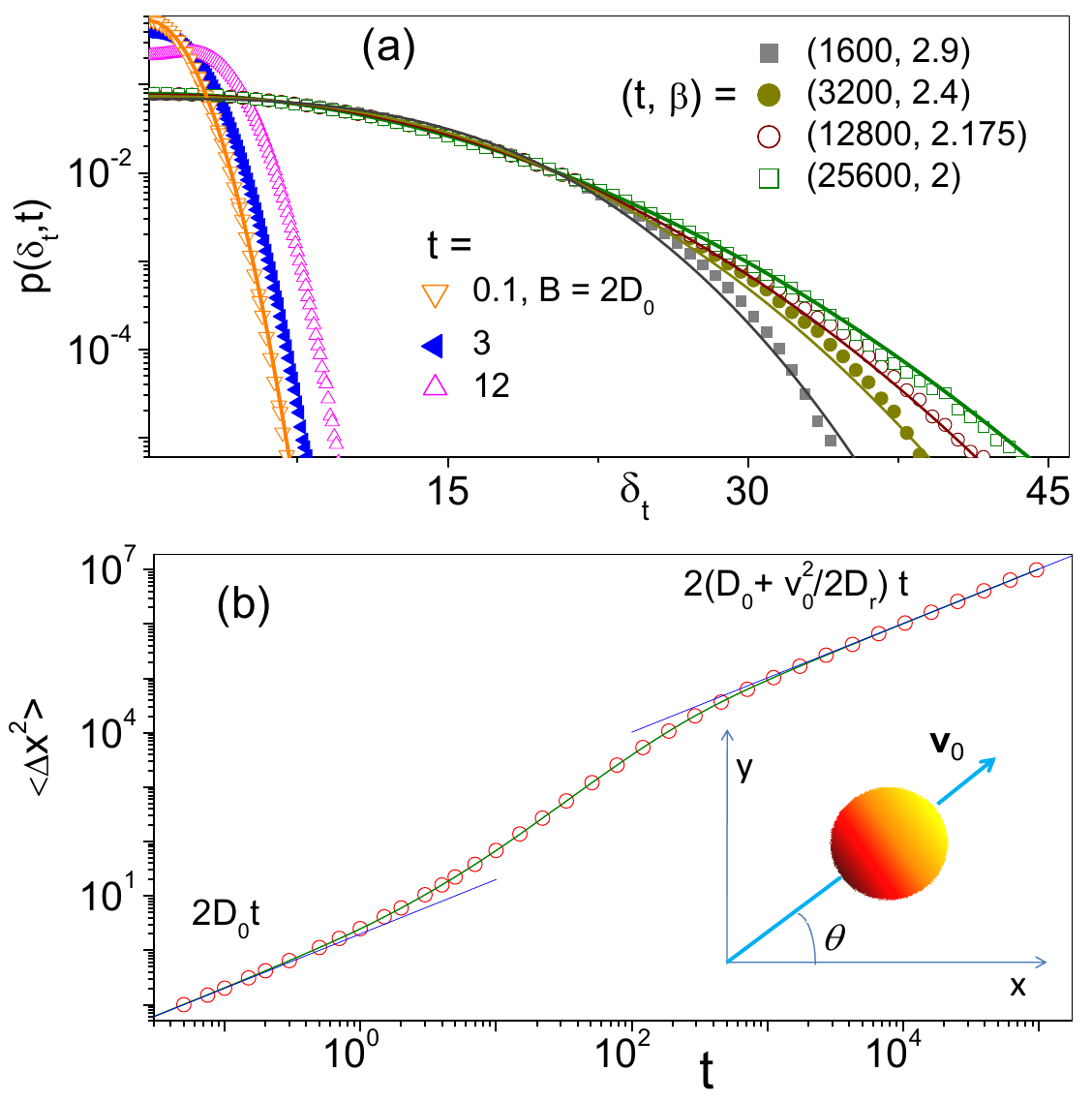}
\caption{Symmetric Janus particle diffusing in a homogeneous medium with
 $D_0=1$, $v_0=1$, $D_\theta=0.01$, and
uniform distributions of the particle's initial position and orientation,
Eqs. (\ref{J1})-(\ref{J2}):  (a) displacement pdf's at different observation
times, $t$; (b) diffusion law, $\langle \Delta x^2\rangle$ vs. $t$.
The numerical data agree well with the analytical law of Eq. (\ref{J3}) (solid curve);
the normal diffusion limits at large and short $t$, Eqs. (\ref{J4}) and (\ref{J5}), are drawn for
a comparison (dashed lines); the pdf's in (a) have been fitted by means of
Eq. (\ref{B3}) with $D$ fitting the large-$t$ data in (b) and an appropriate choice of $\beta$
(see legend). The pdf's with $\beta=2$ at the shortest and largest $t$, panel (a),
are Gaussian curves with half-variance $D_0$ and $D_0+D_s$, respectively.
\label{F2}}
\end{figure}

The overdamped dynamics of a pointlike active JP can be formulated by means
of two translational and one rotational Langevin equation
\begin{eqnarray}
\dot x&=& v_0\cos \theta + \xi_x(t), ~~~ \dot y=v_0 \sin\theta+ \xi_y(t), \label{J1}\\
\dot \theta&=&\xi_\theta (t) \label{J2},
\end{eqnarray}
where $x$ and $y$ are the coordinates of the particle's center of mass, and
the self-propulsion velocity has constant modulus, $v_0$, and orientation
$\theta$, taken with respect to the $x$ axis, see sketch in Fig. \ref{F2}(b).
The translational noises in the $x$ and $y$ directions, $\xi_x(t)$ and
$\xi_y(t))$, and the rotational noise, $\xi_\theta (t)$, are stationary,
independent, delta-correlated Gaussian noises, $\langle
\xi_i(t)\xi_j(0)\rangle = 2 \delta_{ij}D_i\delta (t)$ with $i,j=x,y,\theta$.
The noise strengths $D_x=D_y=D_0$ (isotropic translational fluctuations) and
$D_\theta$ are assumed here to be unrelated for generality (e.g., to account
for different self-propulsion mechanisms \cite{ourPRL}). The reciprocal of
$D_\theta$ is the correlation (or angular persistence) time, $\tau$, of the
self-propulsion velocity. For simplicity, we ignore chiral effects due to
unavoidable fabrication defects \cite{Wang,Loewen,chiral1}. It is worthy
comparing the Langevin Eqs. (\ref{P1})-(\ref{P2}) and (\ref{J1})-(\ref{J2}):
for the ellipsoidal particle anisotropy is geometric, i.e., due to its
elongated shape, whereas, for a pointlike JP anisotropy is dynamical, i.e.,
associated with the instantaneous orientation of its self-propulsion
velocity.

A detailed analytical treatment of the Langevin Eqs. (\ref{J1})-(\ref{J2}) is
to be found in Ref. \cite{Franosch_SciRep}. The unidirectional diffusion of a
free JP in 2D reads \cite{Golestanian07,Loewen09,chiral2},
\begin{eqnarray} \label{J3}
\langle \Delta x^2(t)\rangle=2(D_0+D_s)t+D_s\tau(e^{-|t|/\tau}-1),
\end{eqnarray}
which approaches the Einstein law,
\begin{eqnarray} \label{J4}
\langle \Delta x^2(t)\rangle=2(D_0+D_s)t,
\end{eqnarray}
only for $t\gg \tau$. Here, the unidirectional diffusion constant, $D$,
consists of two distinct contributions, a translational, $D_0$, and a
self-propulsion term, $D_s=v_0^2/2D_\theta$. Instead, for short observation
times Eq. (\ref{J3}) tends to
\begin{eqnarray} \label{J5}
\langle \Delta x^2(t)\rangle=2D_0 t,
\end{eqnarray}
that is, to the normal diffusion law of a passive particle with $v_0=0$. The
analytical law of Eq. (\ref{J3}) and its normal limits for large and small
observation times compare well with our simulation results in Fig.
{\ref{F2}}(b).

The displacement distribution, $p(\delta_t,t)$, exhibits a Gaussian profile
both for $t\to 0$ and $t \to \infty$, but with different half-variances,
respectively $D_0$ and $D=D_0+D_s$, see Fig. \ref{F2}(a). The crossover
between these two Gaussian limits is characterized by platykurtic transient
pdf's with fast decaying tails. Experimental evidence of this phenomenon has
been reported in Ref. \cite{Loewen2}. In the limit $t\to 0$, the displacement
distributions become sensitive to the particle's initial orientation. For a
uniform distribution of $\theta(0)$, shown in Fig. \ref{F2}(a), the rescaled
pdf's approach a Gaussian function with half-variance $D_0$, as to be
expected for an isotropic persistent Brownian motion in the ballistic regime,
$t \ll \tau$. However, for a fixed value of $\theta (0)$, say, $\theta
(0)=0$, the pdf is still a Gaussian with the same half-variance, $D_0$, but
its center moves to higher $\Delta x$ values, with $\langle \Delta x
(t)\rangle=v_0t$ \cite{sperm} (not shown). For intermediate observation
times, $t\simeq \tau$, the displacement pdf's develop two symmetric maxima a
distance of the order of the persistence length, $\Delta x \sim v_0\tau$,
from their centers \cite{Loewen2}.

The different nature of the diffusion transients of ellipsoidal and active
JP's can be easily explained in terms of the coarse grained model of Sec.
\ref{model1}. The orientation of a JP is time correlated; from Eqs.
(\ref{J1})-(\ref{J2}), $\langle \cos \theta (t) \cos \theta (0)
\rangle=\langle \sin \theta (t) \sin \theta (0)
\rangle=(1/2)\exp(-|t|/\tau)$. This implies that both the orientation and the
length of the discrete steps in the $x$ direction, $\Delta x_i$ are time
correlated; given any pair of steps, $\Delta x_i$ and $ \Delta x_j$, both
their time correlations vanish asymptotically only for $|t_j-t_i| \gg \tau$.
However, on comparing panels (a) and (b) of Fig. \ref{F2} we notice the
existence of a rather wide range of observation times, where $\langle \Delta
x^2(t) \rangle$ has approached a linear function of $t$, Eq. (\ref{J4}),
while the rescaled $\Delta x$ distributions are still apparently platykurtic.

The platykurtic nature of this transient is consistent with the coarse
grained model of Sec. \ref{model1}, because the angular correlation of the
self-propulsion velocity vector amounts to an oscillatory behavior of $D_i
-\langle D_i\rangle$, a necessary condition to observe a negative excess
kurtosis, $\mu_x<0$. It remains to explain why, like in Sec. \ref{Perrin},
the onset of normal diffusion anticipates the onset of the Gaussian $\Delta
x$ statistics. We know \cite{Loewen2} that the self-propulsion mechanism of
Eqs. (\ref{J1})-(\ref{J2}) is responsible for the non-Gaussian profile of
$p(\delta_t)$, an effect mitigated by the translational noise as long as
$2D_0t > l_\theta^2$, where $l_\theta = v_0 \tau$ is the JP persistence
length. Therefore, the Gaussian statistics of the unidirectional JP
displacements is expected to emerge only for $t > \tau_*$, with
$\tau_*=\tau(D_s/D_0)$. Note that in the simulations of Fig. \ref{F2} we set
$\tau_*>\tau$.

The results presented in this section lead us to conclude that we are in the
presence of another manifestation of the NGND phenomenon.

\section{Transient Displacement Distributions} \label{beta}

As mentioned in Sec. \ref{intro}, the notion of NGND is commonly associated
with the existence of a wide interval of observation times,  where diffusion
follows a normal law with fixed constant, $D$, and the rescaled displacement
distribution, $p(\delta_t)$, decays (almost) exponentially independently of
$t$. The exponential to Gaussian crossover is hardly accessible to direct
observation \cite{Granick1}. In the low dimensional systems investigated
here, instead, such a transition takes place over a relatively narrower $t$
interval, which led us to look for a phenomenological function
$p(\delta_t,t)$ fitting our simulation data from transient up asymptotic $t$
values.

Contrary to the diffusing diffusivity models, where the limiting Laplace and
Gaussian distributions are functions of the sole diffusion constant, $D$,
 a more realistic fitting procedure needs at least one additional
parameter, $\beta$, to capture the $t$-dependence of the transient pdf's.
Inspired by the numerical findings of Secs. \ref{Perrin} and \ref{Janus}, we
started from the compressed exponential function
\begin{equation} \label{B1}
p(\delta_t)=p_0 e^{-(\delta_t/\delta_0)^\beta},
\end{equation}
where $\beta \geq 1$. The scaling factor, $\delta_0$, and the
 normalization constant, $p_0$, have been computed by imposing the conditions
\begin{equation} \label{B2}
\int_0^\infty p(\delta_t)d\delta_t=1, ~~~\int_0^\infty \delta_t^2p(\delta_t)d\delta_t=2D,
\end{equation}
to obtain the one-parameter {\em ad hoc} fitting function,
\begin{equation} \label{B3}
p_\beta(\delta_t)= \frac{\beta}
{\Gamma(\frac{1}{\beta})^\frac{3}{2}} \left [\frac{\Gamma(\frac{3}{\beta})}{2D} \right ]^\frac{1}{2}
\exp \left[-\left(\frac{\delta_t^2}{2D}
\frac{\Gamma(\frac{3}{\beta})}
{\Gamma(\frac{1}{\beta})}
\right)^{\frac{\beta}{2}} \right].
\end{equation}
This function has been derived phenomenologically starting from the standard
stretched exponential distribution, $p_\beta (\delta_t)=A
\exp(-B\delta_t^\beta)$. The constants $A$ and $B$ have then be determined by
normalizing $p_\beta (\delta_t)$ to one and ensuring that its second moment
yields $\langle \delta_t^2\rangle=2D$ for any value of the free parameter
$\beta$. In view of its derivation, the heuristic distribution (\ref{B3}) may
apply also to the transients of microscopically non-Gaussian diffusion models
\cite{nonG1,nonG2,nonG3}. The fitting parameter $\beta$ is allowed to vary
with $t$; it assumes values in the range $1 \leq \beta \leq 2$ for
leptokurtic distributions (positive excess kurtosis) and $\beta \geq 2$ for
platykurtic distributions (negative excess kurtosis).

The fits of the pdf's drawn in panels (a),(c) of Figs. \ref{F1} and (a) of
Fig. \ref{F2} have been generated from Eq. (\ref{B3}) by setting $D$ equal to
the diffusion constants that best fitted the large-$t$ diffusion data in the
respective panels (b) and, then, computing $\beta$ to get the best fit of the
rescaled displacement distributions at different $t$. The same fitting
procedure has been applied in Figs. \ref{F3} and \ref{F4} of the forthcoming
sections.

Our phenomenological formula (\ref{B3}) fits rather closely the numerical
pdf's reported in Secs. \ref{Perrin} and \ref{Janus}, at least for
sufficiently large observation times. As a matter of fact, the heuristic
argument leading to the fitting function $p_\beta(\delta_t)$ assumes normal
diffusion at any $t$. This is consistent with the diffusive dynamics of the
ellipsoidal Brownian particle with isotropic i.c., displayed in Figs.
\ref{F1}(a)-(b). However, this cannot be the case, for instance, of the
active JP of Fig. \ref{F2}, whose diffusion law for $t < \tau$ clearly
deviates from the asymptotic law of Eq. (\ref{J4}).  A comparison with the
simulation output confirms that the proposed fitting procedure works well for
both systems in the transient regime, $t > \tau$.

\section{Diffusion in a Time Modulated Channel} \label{channel}

In most numerical and experimental investigations
\cite{Granick1,Granick2,Bhatta,Sung1,Sung2,Granick3} the {\it transient}
distributions of $\delta_t$ are presented as sort of universal functions,
$p(\delta_t)$, which decay with (almost) exponential law independently of
$t$. Sometimes the transient interval is so wide that the
exponential-Gaussian crossover is not accessible to direct observation. The
question then rises as to what extent the low dimensional systems addressed
in this work may share that property. In the notation of Sec. \ref{beta},
this corresponds to determining conditions for the fitting parameter $\beta$
to be constant with $\beta \neq 2$ (non-Gaussian transient) over a wide range
of $t$. Note that in the models of Secs. \ref{Perrin} and \ref{Janus} the
onset of the normal diffusion and the Gaussian statistics regimes, which
delimit the NGND transient, are governed by the sole angular relaxation time
$\tau$ ($\tau_*$ being proportional to $\tau$). In the standard formalism of
the central limit theorem this would correspond to saying that the higher
cumulants of the displacement distribution vanish slower with the observation
time than the second moment approaches its linear growth \cite{Feller}. In
this regard, more interesting are systems where the NGND transients are
delimited by two distinct time constants.

\begin{figure}
\centering \includegraphics[width=8.5cm]{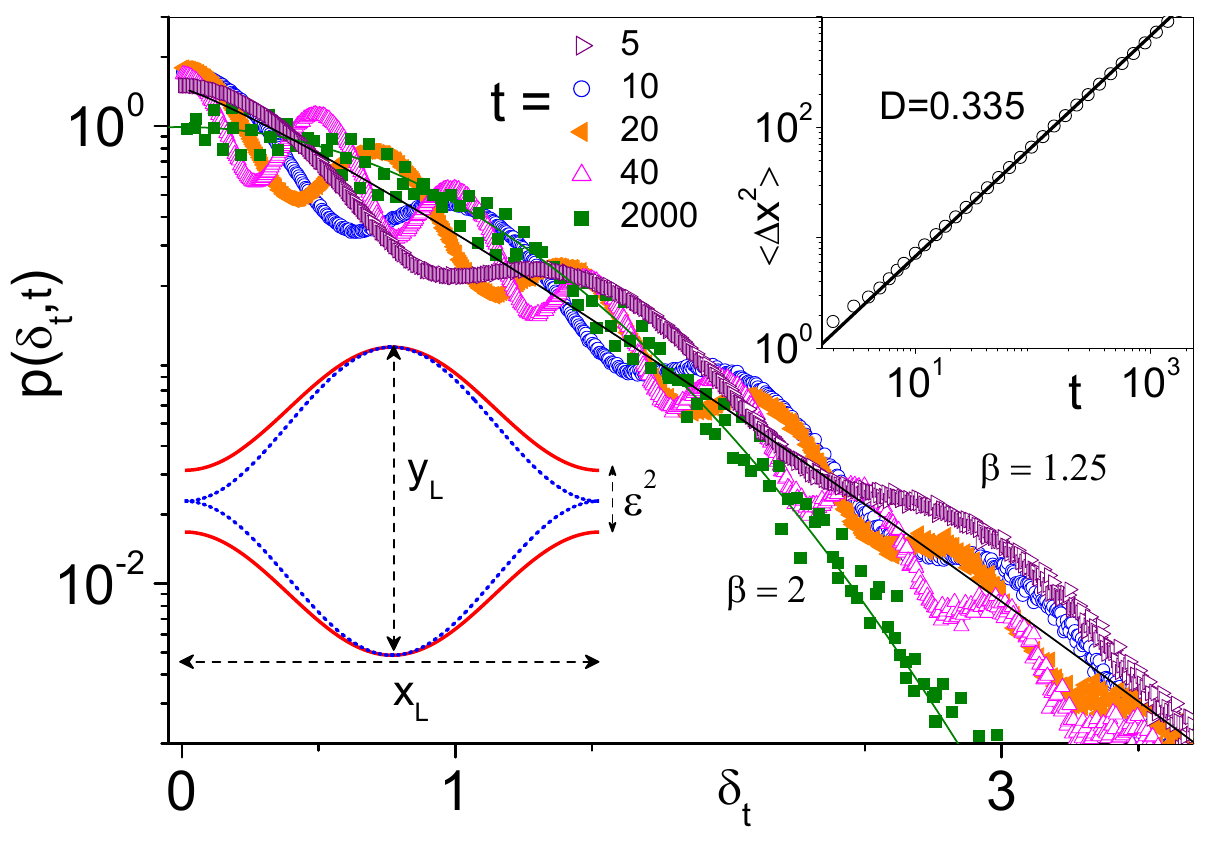}
\caption{Diffusion of a pointlike overdamped particle in a randomly fluctuating channel,
Eq. (\ref{FC2}) with $\varepsilon (t)$ representing the Ornstein-Uhlenbeck process of Eq. (\ref{FC3}).
Simulation parameters are: $y_L=1$, $x_L=\pi$, $D_0=1$,
$D_{\varepsilon}=3$, %$\varepsilon_0=0.1$,
$\tau=50$, and random $x(0)$, $y(0)$ and $\varepsilon (0)$.
In the main panel, rescaled displacement pdf's are shown for increasing observation times, $t$,
in the normal diffusion regime, see data for $\langle \Delta x^2\rangle$ vs. $t$ in the inset.
The fitting $\beta$ values have been obtained from Eq. (\ref{B3}) with $D=0.335$.
At short $t$, the statistics of our data is not good enough to resolve the $t$ dependence
of $\beta$.
\label{F3}}
\end{figure}

A study case is represented by the diffusion of a standard Brownian particle
in a confined geometry \cite{chemphyschem}, the simplest example being a
chainlike structure of cavities connected by narrow pores \cite{PRR}. In 2D,
the dynamics of an overdamped symmetric Brownian particle in a channel is
modeled by two simple Langevin equations
\begin{equation}
\dot x= \xi_x(t), ~~~ \dot y=\xi_y(t), \label{FC1}
\end{equation}
where $x$ and $y$ are the coordinates of the particle's center of mass and
the translational fluctuations $\xi_x(t)$ and  $\xi_y(t)$ are zero-mean,
white Gaussian noises with autocorrelation functions $\langle \xi_i(t)
\xi_j(0)\rangle = 2\delta_{i,j} D_0 \delta (t)$ and  $i,j = x,y$. The
strength of $\xi_i(t)$ coincides with the free-particle diffusion constant,
$D_0$, which is typically proportional to the temperature of the suspension
fluid. However, contrary to Secs. \ref{Perrin} and \ref{Janus}, the particle
is now confined to diffuse inside a narrow corrugated channel with axis
oriented along $x$ and symmetric walls, $y=\pm w(x,t)$. Following Refs.
\cite{chemphyschem}, we assumed for simplicity the sinusoidally modulated
channel half-width
\begin{equation}\label{FC2}
  w(x,t)=(y_L/2)[\varepsilon^2+(1-\varepsilon^2)\sin^2(\pi x/x_L)].
\end{equation}
Here $y_L$ and $x_L$ are respectively the maximum width and the length of the
unit channel cell, sketched in Fig. \ref{F3}, and $\varepsilon^2 y_L$ is the
fluctuating width of the pores located at $x {\rm mod}(x_L)=0$. In the case
of a pointlike particle, hydrodynamic effects \cite{PNAS} can be ignored.
Moreover,  let the width of the channel pores  be time modulated without
affecting the particle's free diffusion constant, $D_0$, for instance, by
applying a tunable external gating potential. Therefore, when integrating the
Langevin Eq. (\ref{FC1}), we neglected the particle radius with respect to
$x_L$ and $y_L$ (pointlike particle approximation) and imposed reflecting
boundary conditions at the walls \cite{ourPRL}.

In Ref. \cite{PRR} we considered the case when $\varepsilon(t)$ is an
Ornstein-Uhlenbeck process
\begin{equation}\label{FC3}
\dot \varepsilon=-\varepsilon/\tau +\sqrt{D_\varepsilon /\tau^2}~\xi_{\varepsilon} (t),
\end{equation}
where $\xi_{\varepsilon}(t)$ is another Gaussian zero-mean valued noise,
independent of $\xi_x(t)$ and $\xi_y(t)$ and delta-correlated, $\langle
\xi_{\varepsilon}(t)\xi_{\varepsilon}(0)\rangle=2 \delta(t)$. The channel
pores open and close randomly in time with average width $\langle
\varepsilon^2 \rangle y_L$, where $\langle \varepsilon^2 \rangle$ coincides
with the variance of $\varepsilon(t)$, $D_{\varepsilon}/\tau = (\pi/2)\langle
|\varepsilon| \rangle^2$.

This channel model manifests prominent NGND, as illustrated in Fig. \ref{F3}.
The displacement distributions are Gaussian for both very short (not shown,
see Ref. \cite{PRR}) and asymptotically long observation times. Indeed, the
particle diffuses freely with constant $D_0$ inside each channel's cell for
$t < \tau_L$, with $\tau_L=x_L^2/8D_0$, before escaping into an
adjacent cell after a mean exit time $\tau_0=\tau_L/\langle
|\varepsilon|\rangle$ \cite{PRR}. For $t \gg \tau_0$, the $x$ directed
diffusion process can thus be described as a random walker with spatial step
$x_L$ and time constant $\tau_0$; memory of the i.c. adopted in our
simulations is completely erased. The ensuing mean square displacement then
follows the Einstein law with approximated diffusion constant
$D=x_L^2/2\tau_0$ \cite{JCP137}.  The displacement distribution assumes its
asymptotic Gaussian profile only for observation times much larger than the
correlation time of the pore fluctuations, $t \gg \tau$. In the simulations
of Fig. \ref{F3}, we set $\tau \gg \tau_0$, which thus defines a NGND
transient interval, $(\tau_0, \tau)$, where diffusion is normal, but the
displacement distributions are non-Gaussian. By taking such interval wide
enough, the transient pdf's, $p(\delta_t,t)$, grow insensitive to $t$, and so
do the fitted $\beta$ values. This way, we mimic the situation reported in
the literature \cite{Granick1,Granick2,Bhatta,Sung1,Sung2,Granick3} for more
complex systems.

This prescription for NGND control is independent of the detailed statistics
of the pore fluctuations. For instance, one can consider the case of a
periodically time modulated pore width with
\begin{equation}\label{FC4}
\varepsilon(t)=\delta_\varepsilon \cos(t/\tau).
\end{equation}
\begin{figure}[t!]
\centering \includegraphics[width=8.5cm]{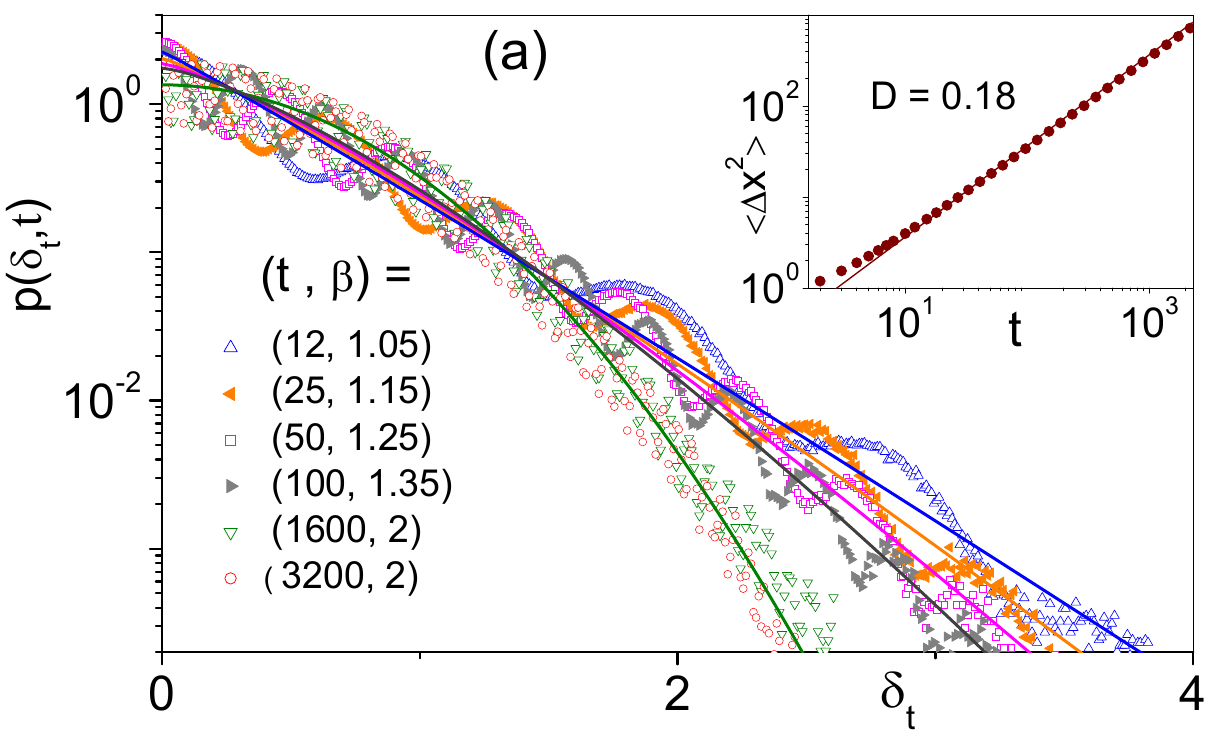}
\centering \includegraphics[width=8.5cm]{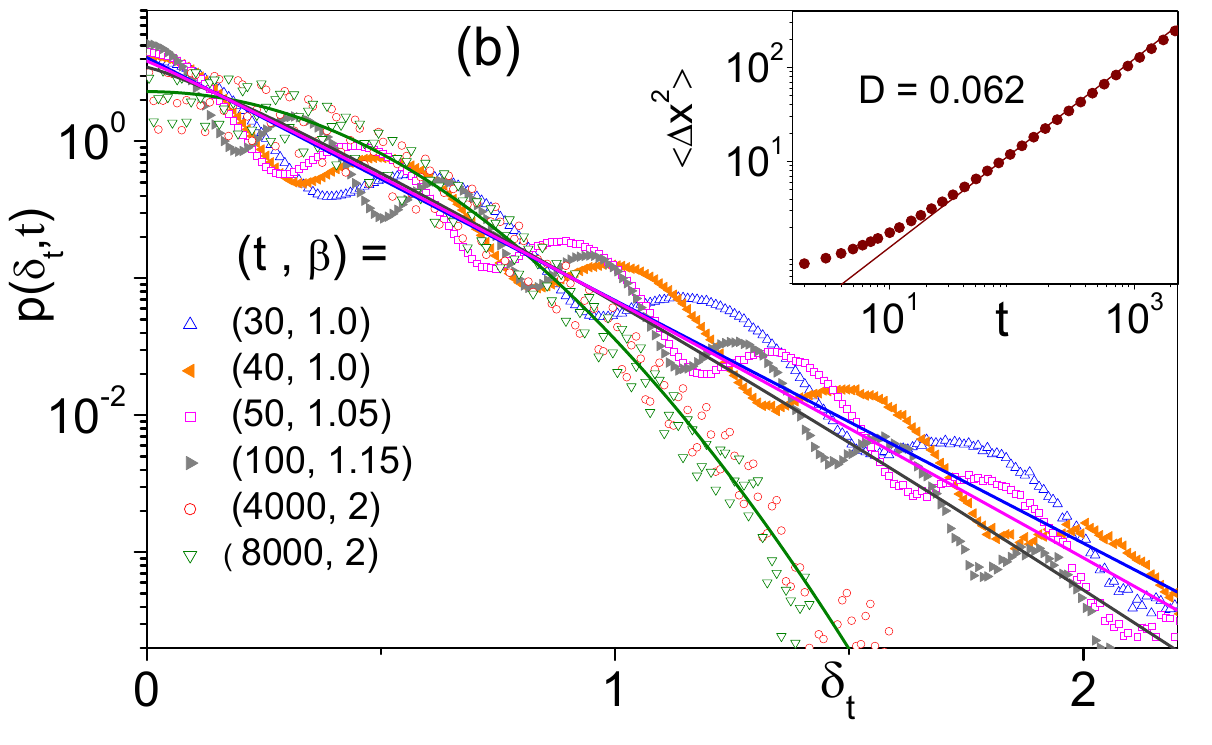}
\caption{Diffusion of a pointlike overdamped particle in a corrugated channel,
Eq. (\ref{FC2}), with $\varepsilon (t)$ representing the periodically time modulation of Eq. (\ref{FC4}).
Simulation parameters are: $y_L=1$, $x_L=\pi$, $D_0=1$,
$\delta^2_{\varepsilon}=0.03$, and (a) $\tau=100$, (b) $\tau=500$. The initial conditions were set by
imposing uniform distributions of the particle's initial position and $\varepsilon (t)$ initial phase.
In the main panels, rescaled displacement pdf's are plotted for increasing $t$,
in the normal diffusion regime. The values of $\beta$ in the legend have been obtained
by fitting Eq. (\ref{B3}) with $D$ computed numerically from the asymptotes, $\langle \Delta x^2\rangle=2Dt$,
drawn in the insets. %The data statistics suffices to resolve the $t$-dependence
%of the fitting parameter $\beta$ only in panel (a).
\label{F4}}
\end{figure}
The simulation data plotted in Fig. \ref{F4}(a) confirm the existence of the
NGND transient interval $(\tau_0, \tau)$, where $\tau$ is now the period of
the sinusoidal function $\varepsilon (t)$ of Eq. (\ref{FC4}) and $\langle
\varepsilon^2\rangle= \delta_\varepsilon^2/2=(\pi^2/8)\langle
|\varepsilon|\rangle^2$.  For a quantitative comparison, in Figs. \ref{F3}
and \ref{F4}(a) $\langle \varepsilon^2\rangle$ have been assigned the same
value, so that for the simulation parameters of Fig. \ref{F4} both time
scales, $\tau_0$ and $\tau$, are larger by approximately the same factor two.

The NGND phenomenon in Fig. \ref{F4}(a) is apparent. The rescaled pdf's shown
there are clearly non-Gaussian. Fat oscillating tails arise for $t > \tau$, as an effect of the spatial periodicity of the channel. Indeed, the oscillation period of the plotted distributions is of the order
$x_L/\sqrt{t}$. This effect, detectable also in Figs. \ref{F3} and
\ref{F4}(b), plays here a marginal role. Indeed, on disregarding such
oscillations, the tails of the non-Gaussian distributions can still be fitted
by the function of Eq. (\ref{B3}) for an appropriate choice of the free
parameter $\beta$.

However, in Fig. \ref{F4}(a) and in contrast with Fig. \ref{F3},  the
deterministic nature of the pore time modulation allowed us to resolve a weak
$t$-dependence of $\beta$, without increasing the statistical accuracy of our
simulation runs.
It is important to remark that such a residual $t$-dependence of the
transient rescaled pdf's can be further suppressed by widening the transient
interval $(\tau_0, \tau)$. An example is shown in panel (b) of Fig. \ref{F4},
where the simulation parameters are the same as in panel (a), except for the
modulation period, $\tau$, which is five times larger. We remark that, for
the simple model at hand, this result is analytically predictable upon
reformulating the particle's dynamics in the dimensionless units, $x \to
x/x_L$, $y \to y/x_L$ and $t \to t/\tau$.

\section{Diffusion in Convection Rolls} \label{Pomeau}

We finally address the reasons why transients under NGND conditions can be
either lepto- or platykurtic. In Secs. \ref{Perrin} and \ref{Janus} we looked
at two simple models, which exhibit distributions of the one or the other
type, respectively, with $1\leq \beta \leq 2$ and $\beta \geq 2$. We consider
now a slightly more complicated 2D system, which can undergo both transients,
depending on the choice of its dynamical parameters. The numerical analysis
of its diffusion properties will help us shed light on the different
microscopic mechanisms responsible for these two type of transients, thus
justifying the generalization of the NGND notion proposed in this work.

\begin{figure}[t!]
\centering \includegraphics[width=8.0cm]{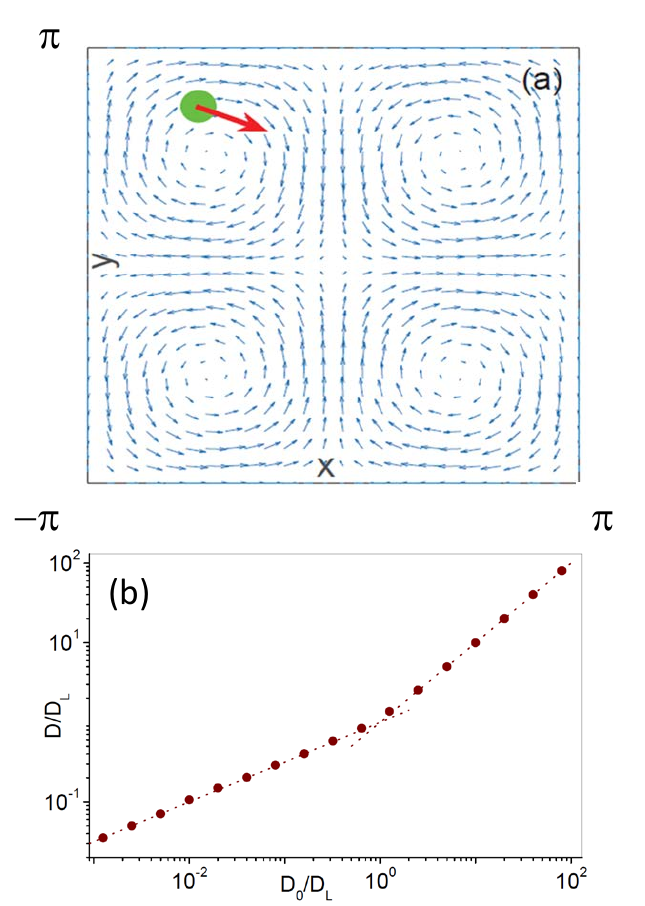}
\caption{Diffusion in the periodic convective
flow pattern of Eq. (\ref{Po1}): (a) Flow cell unit consisting of four
counter-rotating subcells; (b) The asymptotic diffusion  constant, $D$ vs.
$D_0$: the numerical data (dots) are compared with the analytical prediction
discussed in the text, see Eq. (\ref{Po3}). The stream function parameters
are $U_0=1$ and $L=2\pi$, and the diffusion scale is $D_L=U_0L/2\pi$.
\label{F5}}
\end{figure}

To this purpose we investigated the diffusion of a pointlike overdamped
particle of coordinates $x$ and $y$, suspended in a stationary planar laminar
flow with periodic center-symmetric stream function
\cite{Gollub1,Gollub2,Rosen,Pomeau,Saintillan,Vulpiani,Neufeld}.
\begin{equation}
\label{Po1}
\psi(x,y)= ({U_0L}/{2\pi})\sin({2\pi x}/{L})\sin({2\pi y}/{L}),
\end{equation}
where $U_0$ is the maximum advection speed and $L$ the wavelength of the flow
unit cell. The ensuing particle's dynamics can be formulated in terms of two
driven Langevin equations,
\begin{eqnarray} \label{Po2}
\dot {x}= u_x + \xi_x(t), ~~~~ \dot {y}= u_y + \xi_y(t),
\end{eqnarray}
with the vector $(u_x, u_y) =(\partial_y, -\partial_x)\psi$ representing the
local advection velocity. As illustrated in Fig. \ref{F5}(a), this defines
four counter-rotating flow subcells, also termed convection rolls. The
translational noises, $\xi_i(t)$ with $i=x,y$ are stationary, independent
Gaussian noises with auto-correlation functions $\langle
\xi_i(t)\xi_j(0)\rangle = 2 D_0 \delta_{ij}\delta (t)$. They can be regarded
as modeling homogeneous, isotropic thermal fluctuations. In our simulations,
the flow parameters, $U_0$ and $L$ were kept fixed, as they define the
natural length and time units, $L$ and $\Omega_L^{-1}=L/2\pi U_0$, respectively. Therefore,
the only tunable parameter left is the noise strength, $D_0$. Having in mind
a stationary system, we assumed uniform distributions of the initial
particle's coordinates, $x(0)$ and $y(0)$. Indeed, due to the
incompressibility of $(u_x,u_y)$, in the presence of translational noise, a
particle's trajectory eventually fills up the flow unit cell uniformly. To
this regard, we remind that, in the absence of noise, the advection period
tends to diverge as the closed trajectory of a dragged particle runs close
the subcell boundaries \cite{Weiss}; hence, for $D_0=0$ the particle gets
trapped in a convection roll \cite{Rosen,Pomeau,Neufeld}.
\begin{figure}
\centering \includegraphics[width=8.5cm]{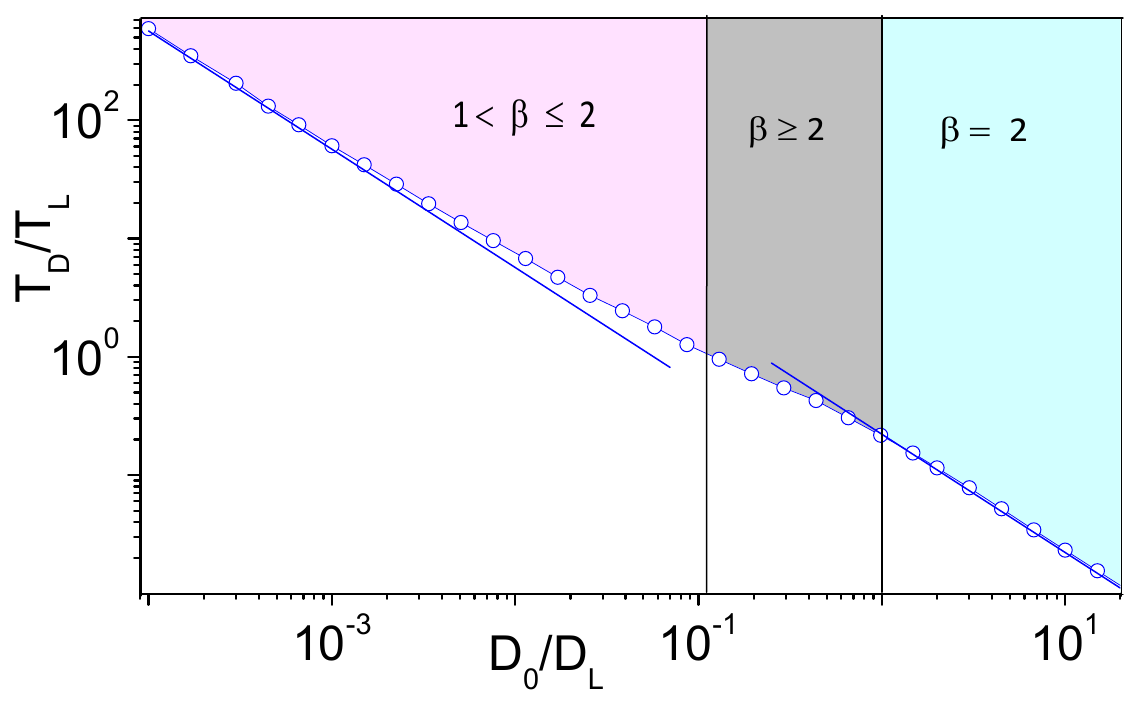}
\caption{Diffusion mechanisms in the periodic flow pattern
of Eq. (\ref{Po1}): mean first-exit time, $T_D$, vs. thermal
noise, $D_0$. Convection flow parameters are $U_0=1$ and $L=2\pi$.
The asymptotic solid lines on the left and right are respectively
$T_0$ and $T_0/4$, with $T_0$ given in Eq. (\ref{Po4}); the horizontal
dashed line represents the advection period, $T_L$. Three $D_0$ intervals.
shaded in different colors, are separated by $D_*$, obtained by
imposing $T_D=T_L$, and $D_L$, defined in Fig. \ref{F5}. In each interval,
the range of $\beta$ is reported for reader's convenience;
no NGND was detected for $D_0>D_L$.
\label{F6}}
\end{figure}
Particle transport in such a flow pattern has been studied under diverse
conditions and a rich phenomenology has emerged \cite{Gollub1,Gollub2,
Rosen,Pomeau,Saintillan,Vulpiani,Neufeld}. We focus here on the Brownian
diffusion of a passive particle under the simultaneous action of
translational fluctuations and advective drag. A first important feature of
this system is illustrated in Fig. \ref{F5}(b), where we plotted the
dependence of the asymptotic diffusion constant, $D$, on the noise intensity
(and particle's no-flow diffusion constant), $D_0$. The mean square
displacement is an asymptotically linear function of time for any choice of
$D_0$. However, on increasing $D_0$, the diffusion constant, $D$, changes
from
\begin{equation} \label{Po3}
D=\kappa \sqrt{D_L D_0},
\end{equation}
for $D_0< D_L$ (advective diffusion), to $D=D_0$, for $D_0>D_L$ (thermal
diffusion), an abrupt crossover occurring at $D_0\simeq D_L$, with $D_L=U_0
L/2\pi$ \cite{RR2}. The constant $\kappa$ depends on the geometry of the flow
cells \cite{Rosen,Pomeau}; for a 2D array of square counter-rotating
convection rolls, $\kappa \simeq 1.06$ \cite{Rosen}. This property can be
explained with the fact that for $D<D_L$ the spatial diffusion occurs along
the separatrices delimiting the four subcells of the stream function,
$\psi(x,y)$, of Fig. \ref{F5}(a). Stated otherwise, the diffusion process is
regulated by the advection velocity field \cite{ourPOF}. Vice versa, for
$D_0>D_L$ the effects of advection on the particle's diffusion become
negligible. Not surprisingly, we detected NGND only for $D_0< D_L$.

The diffusion process is governed by two competing mechanisms: (i) Particle's
circulation inside the counter-rotating subcells of $\psi(x,y)$. The
corresponding vorticity, $\nabla \times {\vec u}= -\nabla^2\psi$, has a
maximum, $\Omega_L=2\pi U_0/L$, at the center of the subcells. This defines
the time scale, $T_L=2\pi/\Omega_L$, for the advection period, that is an
estimate of the average time taken by advection to drag the particle around a
convection roll; (ii) Diffusion across the convection rolls. The mean
first-exit time, $T_D$, of a Brownian particle out of a unit convection cell
of $\psi(x,y)$, can be easily computed for $D_0\gg D_L$ simply by ignoring
advection \cite{Gardiner},
\begin{equation}\label{Po4}
T_0= \frac{1}{D_0} \left (\frac{L}{2\pi} \right )^2\left ( \frac{4}{\pi}\right )^4 ~\sum_{m,n}^{(\rm odd)}\frac{1}{m^2}\frac{1}{n^2}\frac{1}{m^2+n^2},
\end{equation}
where the summation is restricted to the odd values of $m$ and $n$. In the
opposite limit, $D_0\ll D_L$, $T_D$ is just one fourth of $T_0$, because, as
anticipated above, at very low noise levels, the exit process consists of a
slow activation mechanism, which takes the particle from the center of a
subcell to its boundaries, followed by a relatively faster flow-driven
propagation along the grid formed by the subcell separatrices.

\begin{figure}[b!]
\centering \includegraphics[width=8.5cm]{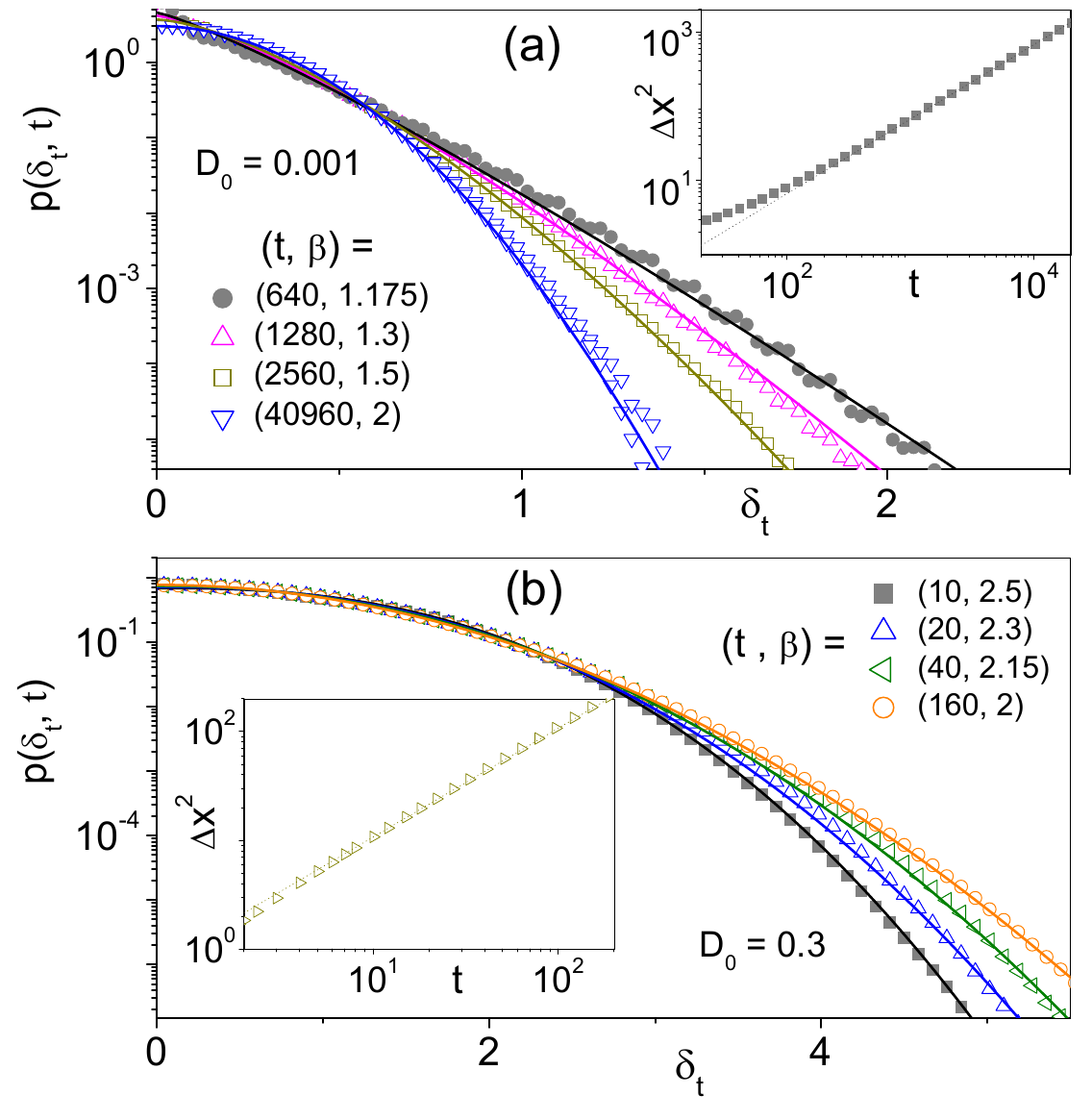}
\caption{Leptokurtic, $D_0<D^*$ (a), and platykurtic transients, $D^* < D_0 < D_L$ (b),
in a periodic array of 2D convection rolls. $D_0$, $t$ and $\beta$ are reported in the
legends; convection flow parameters are $U_0=1$ and $L=2\pi$.
All transient pdf's were taken in the regime of normal diffusion,
see insets. %for numerical data of $\langle \Delta x^2\rangle$ vs. $t$.
\label{F7}}
\end{figure}
Thanks to thermal fluctuations, the Brownian particle jumps from roll to
roll, thus diffusing in the $x,y$ plane. Its coarse-grained motion can be
modeled as a discrete random walker with time constant $T_D$ \cite{Gardiner}.
Therefore, for large observation times, $t > T_D$, the particle executes
normal diffusion.

For the simulation parameters adopted in Fig. \ref{F6}, the crossover between
low- and high-noise estimates of $T$, respectively $T_D=T_0$  and
$T_D=T_0/4$, occurs in the region of advective diffusion, $D_0 <D_L$. More
remarkably, it appears to correspond to the condition, $T_D=T_L$, namely,
when the two competing time scales of the particle's dynamics inside a
convection roll  coincide. Such a condition defines a unique $D_0$ value,
$D_*$, which splits the advective diffusion domain into two distinct
subdomains, respectively, $D_0<D_*$ and $D_*<D_0<D_L$.

Numerical simulation clearly shows evidence of NGND for $t > T_D$, in
close analogy with the models of Secs. \ref{Perrin} and \ref{Janus}, except
for an important peculiarity: The transient displacement distributions
displayed in Fig. \ref{F7}, turn out to be leptokurtic, with $1 \leq \beta
\leq 2$, for $D_0 < D_*$, and platykurtic, with $\beta \geq 2$, for $D_* <
D_0 < D_L$. This can be explained with the fact that here the displacement
length correlation, see Sec. \ref{model1}, is dominated by thermal noise in
the lower $D_0$ interval, where $T_D>T_L$, and by advection in the larger
$D_0$ interval, where $T_L>T_D$. Accordingly, in the formulation of Secs.
\ref{Perrin} and \ref{Janus}, the role of transient time, $\tau$, is played
respectively by $T_D$ for $D_0<D_*$ and by $T_L$ for $D_0 > D_*$.

For $D_* \ll D_0 < D_L$, the onset of normal diffusion and the
exponential-Gaussian transitions are thus regulated by two distinct time
scales, respectively, $T_0$ and $T_L$. Indeed, the slowest time modulation of
the particle's dynamics is due to the advective drag inside the convection
rolls. By generalizing our discussion for the NGND of a free JP, Sec.
\ref{Janus}, we  conclude that such a rotational dynamics must be responsible
for the negative excess kurtosis of the unidirectional particle's
displacements reported in Fig. \ref{F7}(b). The range of the $\beta$ values,
fitted according to the procedure of Sec. \ref{beta}, is shown in Fig.
\ref{F6} for each $D_0$ interval.

In conclusion, the NGND transients of this model can change from leptokurtic
to platykurtic simply by raising the strength of the internal noise. Most
remarkably, this and related diffusive systems are easily accessible to
direct experimental demonstration \cite{Gollub2,Saintillan}.

\section{Conclusions} \label{conclusions}

In this work we have investigated NGND transients
\cite{Granick1,Granick2,Bhatta,Sung1,Sung2,Granick3} in low dimensional
stochastic processes. These become apparent when the Einstein law, which
characterizes normal diffusion, sets in for observation times, $t$, shorter
than the asymptotic Gaussian displacement statistics, predicted by the
central limit theorem. A wide class of
 low dimensional systems manifest NGND under the condition that their
local dynamics is subjected to time correlated modulations.

Time modulation can affect the effective particle geometry (e.g., its
cross-section in the diffusion direction, Sec. \ref{Perrin}), its dynamics
(e.g., its isotropic self-propulsion mechanism, Sec. \ref{Janus}), or its
confinement geometry (e.g., the cross-section of the directed channel
containing the particle, Sec. \ref{channel}). In all cases discussed here the
system's modulation is time correlated with time constant, $\tau$, larger
than any other microscopic dynamical time scale. We then noticed that NGND
becomes more prominent when the onset times of normal diffusion and the
Gaussian displacement statistics are well separated, with the former much
lower than the latter. This situation is well illustrated by the fluctuating
narrow channel of Sec. \ref{channel}, where normal diffusion occurs for $t$
larger than the mean pore crossing time and the Gaussian statistics sets in
for $t$ larger that the tunable correlation time of the pore modulation.

In low dimensional systems, NGND features exhibit a smooth dependence on the
observation time. The transient rescaled displacement distributions are not
``universal'' over large $t$ intervals, in sharp contrast with the extended
disordered systems first studied in the literature
\cite{Granick1,Granick2,Bhatta,Sung1,Sung2,Granick3}. To quantify the
$t$-dependence of the transient pdf's we introduced an {\it ad hoc} fitting
function, $p_\beta(\delta_t)$, which, by construction, reproduces the normal
diffusion law, with diffusion constant obtained by direct observation, and
fits the numerical curves $p(\delta_t,t)$ by tuning only one free parameter,
$\beta$. Actually, in Sec. \ref{channel} we noticed that by increasing the
gap between the two distinct time scales defining the transient interval, the
$t$-dependence of $\beta$ is suppressed, with $\beta$ tending to one (Laplace
distribution). This situation closer resembles the current description of the
NGND phenomenon in complex systems.

However, NGND in low dimensional systems has the advantage of being easily
controllable by tuning the time modulation of the microscopic dynamics. For
instance, the two simplest models discussed here, the free ellipsoidal and
Janus particles, exhibit remarkably different transient distributions,
respectively with fat, $1 \leq \beta \leq 2$, and thin tails, $\beta \geq 2$.
Platykurtic transient distributions are peculiar to systems with rotational
modulation of the diffusion process, because, as discussed in Sec.
\ref{model1}, this can cause negative time correlations of the unidirectional
displacement lengths; hence the negative values of the excess kurtosis.

This conclusion is corroborated by Brownian diffusion in the periodic array
of 2D convection rolls discussed in Sec. \ref{Pomeau}. In contrast with the
elementary models of Secs. \ref{Perrin} and \ref{Janus}, there the physical
mechanism determining the transient time varies depending on the strength of
the thermal fluctuations. At low temperatures, the transient dynamics of the
particle is governed by isotropic random jumps from convection roll to
convection roll, largely insensitive to the details of its trajectory inside
the individual rolls. On the contrary, at higher temperatures, but still in
the advective diffusion regime, roll jumping grows faster compared with the
circulation inside the rolls; transients are then dominated by a rotational
dynamics, which causes a negative excess kurtosis of the particle's
displacements.

We conclude now mentioning a number of open issues we intend to address in
the next future.

(i) We showed that low-dimensional systems exhibit NGND transients for
observation times not too much larger than their largest intrinsic relaxation
time. It remains to be seen how one can make such transient time intervals
wider, for instance, by means of a hierarchy of additional stochastic degrees
of freedom.

(ii) We wonder to what extent our discussion of discrete NGND in Sec.
\ref{model1} is related to the formalism of the large deviations theory
\cite{LDT}. This might provide an alternate phenomenological description of
the NGND transients, also applicable to higher dimensional systems.

(iii) NGND transients in laminar flows are of great relevance in microfluidics. 
This results reviewed in Sec.7 will be published in a more detailed report to appear soon \cite{JFM}.
 We showed that leptokurtic (platykutic) transients are an
effect of the mostly thermal (advective) tracer's diffusion. The question
then rises as how this explanation translates in the cases of turbulent
flows, a recurrent problem in biological systems.

(iv) Finally, it is conceivable that persistent NGND transients impact how
active micro-swimmers interact with each other or with confining walls or
other obstacles, to form all kinds of clustered structures. The implications
of such a mechanism in the technology of active matter need further
investigation.

\acknowledgements{Y.L. is supported by the NSF China under grant No. 11875201 and No. 11935010.
% F.M. is supported under the 1000 Talent Program of China.
P.K.G. is supported by SERB
Start-up Research Grant (Young Scientist) No. YSS/2014/000853 and the UGC-BSR
Start-Up Grant No. F.30-92/2015.}

%%%%%%%%%%%%%%%%%%%%%%%%%%%%%%%%%%%%参考文献的排法%%%%%%%%%%%%%%%%%%%%%%%%%%%%%%%%%%%%
%%%%%%%%%%%%%%%%%%%%%%%%%%%%%%%%%%%%注：\href{}{}%%%第一个括号表示需要链接的网址,第二个花括号表示锭接网址的内容%%%%%%%%%%%%%%%%

%
\end{document}